\RequirePackage{fix-cm}
\documentclass[11pt,a4paper]{article}
\usepackage{graphicx}
\usepackage{mathptmx}      
\usepackage[T1]{fontenc}
\usepackage[utf8]{inputenc}
\usepackage[hidelinks,colorlinks=true,linkcolor=blue]{hyperref}
\hypersetup{citecolor=blue}
\usepackage[all]{hypcap}
\usepackage{xcolor}
\usepackage{mathrsfs,amsmath}
\usepackage[font={small,it}]{caption}

\usepackage{sectsty}
\sectionfont{\large}
\subsectionfont{\normalsize}
\subsubsectionfont{\normalsize}
\paragraphfont{\normalsize}

\begin{document}
\title{Modeling and simulation of microbial enhanced oil recovery including interfacial area}
\author{David Landa-Marb\'an\footnote{Department of Mathematics, University of Bergen, Bergen, Norway (\href{mailto:david.marban@uib.no}{david.marban@uib.no}, \href{mailto:florin.radu@uib.no}{florin.radu@uib.no}, \href{mailto:jan.nordbotten@math.uib.no}{jan.nordbotten@math.uib.no})}, Florin A. Radu${}^\dagger$ and Jan M. Nordbotten${}^\dagger$}
\date{}
\maketitle
\begin{abstract}
\noindent The focus of this paper is the derivation of a non-standard model for microbial enhanced oil recovery (MEOR) that includes the interfacial area (IFA) between the oil and water. We consider the continuity equations for water and oil, a balance equation for the oil-water interface and advective-dispersive transport equations for bacteria, nutrients and surfactants. Surfactants lower the interfacial tension (IFT), which improves the oil recovery. Therefore, we include in the model parameterizations of the IFT reduction and residual oil saturation as a function of the surfactant concentration. We consider for the first time in context of MEOR, the role of IFA in enhanced oil recovery (EOR). The motivation to include the IFA in the model is to reduce the hysteresis in the capillary pressure relationship, include the effects of observed bacteria migration towards the oil-water interface and biological production of surfactants at the oil-water interface. An efficient and robust linearization scheme was implemented, in which we use an implicit scheme that considers a linear approximation of the capillary pressure gradient, resulting in an efficient and stable scheme. A comprehensive, 2D implementation based on two-point flux approximation (TPFA) has been achieved. Illustrative numerical simulations are presented. We give an explanation of the differences in the oil recovery profiles obtained when we consider the IFA and MEOR effects. The model can also be used to design new experiments in order to gain a better understanding and optimization of MEOR.\\[15pt]
\end{abstract}
\newpage
\small
\begin{tabular}{l l}
\textbf{List of Symbols}\\
$A$ & Cross sectional area\\
$a_{ow}$ & Specific IFA\\
$C_b,\;C_n,\;C_s$ & Bacterial, nutrient and surfactant concentration\\
$C^*_n$ & Critical nutrient concentration for metabolism\\
$d_1$ & Bacterial decay rate coefficient\\
$D_{b}^{\text{eff}},\;D_{n}^{\text{eff}},\;D_{s}^{\text{eff}}$ & Effective diffusion coefficients\\
$E_{ow}$ & Production rate of specific IFA\\
$e_{ow}$ & Strength of change of specific IFA\\
$F$ & Source/sink term\\
$\textbf{g}$ & Gravity\\
$g_{1,\text{max}}$ & Maximum bacterial growth rate coefficient\\
$\textbf{k},\;\textbf{k}_o,\;\textbf{k}_w$ & Absolute, oil and water effective permeabilities\\
$k_{a}$ & Diffusion coefficient for the chemotaxis\\
$k_{r,o},\;k_{r,w}$ & Oil and water relative permeabilities\\
$\textbf{k}_{ow}$ & Interfacial permeability\\
$K_{b/n},\;K_{s/n}$ & Half saturation constants for producing bacteria and surfactants\\
$K_{a}$ & Half saturation constant for producing surfactants\\
$l_1,\;l_2,\;l_3$ & Fitting parameters for modeling the reduction of IFT\\
$L$ & Length of porous medium\\
$N_\text{B}$ & Bond number\\
$N_\text{Ca}$ & Capillary number\\ 
$N_\text{T}$ & Trapping number\\
$p,\;p_o,\;p_w$ & Average, oil and water pressure\\
$pv$ & Number of pore volume injected\\
$Q_T$ & Water injection rate\\
$s_o,\;s_w$ & Oil and water saturation\\
$s_{or},\;s^{\text{min}}_{or},\;s^{\text{max}}_{or}$ & Residual oil saturation, minimum and maximum\\
$s_{wi}$ & Irreducible water saturation\\
$s_w^*$ & Effective water saturation\\
$T_1,\;T_2$ & Fitting parameters for modeling the residual oil saturation\\
$\textbf{u},\;\textbf{u}_{\Sigma}$ & Volumetric and total flow rate per area\\
$v_g$ & Settling velocity of bacteria\\
$v_{ow}$ & Interfacial velocity\\ 
$V_p$ & Pore volume\\
$W$ & Width of porous medium\\
$Y_{s/b},\;Y_{s/n}$ & Surfactant yield coefficients per unit bacteria and nutrient\\
$\alpha_0$ & Angle of flow relative to the horizontal\\
$\alpha_1,\;\alpha_2,\;\alpha_3,\;\alpha_4$ & Parameters for the IFA relation\\
$\alpha_{b,L},\;\alpha_{n,L},\;\alpha_{s,L}$ & Longitudinal dispersivities\\ 
$\alpha_{b,T},\;\alpha_{n,T},\;\alpha_{s,T}$ & Transverse dispersivities\\
$\delta$ & Dirac delta\\
$\Delta t$ & Time step\\
$\Delta x$ & Space step\\
$\lambda_o,\;\lambda_w$ & Oil and water mobilities\\
$\mu$ & Viscosity\\
$\mu_{s,\text{max}}$ & Maximum specific biomass production rate\\
$\phi$ & Porosity\\
$\rho$ & Density\\
$\sigma,\;\sigma_{\text{min}},\;\sigma_{\text{max}}$ & IFT, minimum and maximum\\
$\theta$ & Contact angle\\
\end{tabular}\\[20pt]
\begin{tabular}{l l}
\textbf{Subscripts/superscripts}\\
$b$ & Bacteria\\
$n$ & Nutrient\\
$s$ & Surfactant\\
$o$ & Oil\\
$w$ & Water
\end{tabular}\\[1pt]
\begin{tabular}{l l}
\textbf{Abbreviations}\\
BE & Backward Euler\\
EOR & Enhanced oil recovery\\
IFA & Interfacial area\\
IFT & Interfacial tension\\
MEOR & Microbial enhanced oil recovery\\
ODE(s) & Ordinary differential equation(s)\\
PDE(s) & Partial differential equation(s)\\
REV & Representative elementary volume\\
TPFA & Two-point flux approximation\\
\end{tabular}
\section{Introduction}\label{intro}
Among the various sources of energy, oil remains as one of the most valuable ones, considering its extensive use in the daily life, such as in the production of gasoline, plastic, etc. After discovering a petroleum reservoir, one can extract about 15-50\% of the oil by using and maintaining the initial pressure in the reservoir through water flooding (first and second phase oil recovery); however, 50-85\% of oil remains in the reservoir after this, so called conventional recovery \cite{Patel:Article:2015}. This is the motivation for developing new extraction techniques in order to recover the most oil possible. One of these EOR techniques consists of adding bacteria to the reservoirs and using their bioproducts and effects to improve the oil production, which is called MEOR. Besides all MEOR experiments \cite{Armstrong:Article:2012, Hommel:Article:2015}, it is worth pointing out that MEOR has been already used successfully in oil reservoirs \cite{Lazar:Article:2007, Patel:Article:2015}.  Nevertheless, the MEOR technology is not yet completely understood and there is a strong need for reliable mathematical models and numerical tools to be used for optimizing MEOR.\par

The bioproducts formed due to microbial activity are acids, biomass, gases, polymers, solvents and surfactants \cite{Sen:Article:2008}. The main purpose of using microbes (bacteria) is to modify the fluid and rock properties in order to enhance the oil recovery. These microbes and the produced surfactants have the advantage to be biodegradable, temperature tolerant, pH-hardy, non-harmful to humans and lower concentrations of them can produce similar results as chemical surfactants \cite{Patel:Article:2015}.\par 

We can describe briefly the model presented as follows: we inject water, bacteria and nutrients to a reservoir. The bacteria consume nutrients and produce more bacteria and surfactants. As time passes, some bacteria die or reproduce. The surfactants reduce the oil-water IFT, allowing the recovery of more oil. The consideration of IFA in the model allows to include the biological production of surfactants at the oil-water interface \cite{Donaldson:Book:1989}, reduces the hysteresis \cite{Hassanizadeh:Article:1993, Pop:Article:2009} and also enables to include that bacteria is mainly living at the oil-water interface \cite{Kosaric:Book:2015}, which is believed to be a very important feature for MEOR.\par

There exist different systems where the IFA is important. For example, another important application of microorganisms is in the soil remediation \cite{Bollag:Article:2008}. Specially, surfactants can increase bioavailability and degradation of soil contaminants, for example petroleum-derived hydrocarbons \cite{Viramontes-Ramos:Article:2010}. Nevertheless, although the general theory for IFA was established \cite{Hassanizadeh:Article:1990, Niessner1:Article:2008, Pop:Article:2009}, the development of IFA based models for particular applications remains a current challenge. In this work we will derive for the first time a mathematical model for MEOR which includes IFA. It is worth to be mentioned that further developments of the present model, which are considering a formal upscaling from pore to core and in this way better describe the evolution of the micro scale are possible but beyond the aim of this study \cite{van:Article:2010, Bringedal:Article:2016, Musuuza:Article:2009}.\par

Mathematical models for MEOR are based on coupled nonlinear partial differential equations (PDEs) and ordinary differential equations (ODEs), which are very difficult to be solved. Therefore, it is necessary to use advanced numerical methods and simulations to predict the behavior on time of the unknowns in this complex system. For example, in \cite{Nielsen:Article:2015} they used a semi-implicit finite difference technique and in \cite{Li:Article:2011} they used Comsol Multiphysics, that is a commercial PDE solver using finite elements together with variable-step back differentiation and Newton method. Even though it is possible to buy commercial software in the petroleum industry for simulation, it is preferable to do the discretization of the equations and write an own code to perform numerical simulations, in order to implement new relations that are not included in the commercial ones.\par

Most of the MEOR models are based on non-realistic simplifications (for example, only one transport equation for the bacteria is considered, hysteresis in the capillary pressure is neglected, the oil-water interfacial area is not included, numerical simulations are just made in 1D \cite{Kim:Article:2006, Nielsen:Article:2015, Li:Article:2011}). In this general context, the objective of the research reported in the present article was to develop and implement (in a 2D porous media) an accurate numerical simulator for MEOR.\par

To summarize, the new contributions of this paper are
\begin{itemize}
\item the development of a multidimensional comprehensive mathematical model for\\ MEOR, which includes bacteria, nutrients, surfactants and two-phase flow. 
\item the inclusion of the role of IFA in MEOR.
\item the inclusion of the tendency of bacteria to move to the oil-water interface. 
\item the inclusion of the biological production of surfactants at the oil-water interface.
\end{itemize}

The paper is structured as follows
\begin{itemize}
\item Reservoir modeling. We introduce the basic concepts, ideas and equations for modeling MEOR. In addition, we explain the new phenomena we can model in MEOR when we include the IFA. 
\item Discretization and implementation. We explain the techniques we used for the discretization, namely finite differences and TPFA for the spatial discretization and backward Euler (BE) for the time discretization. We also describe the algorithm we used for numerically solving the mathematical model for MEOR.
\item Results and discussion. We present the results of the numerical experiments by studying the effects of the new relations we proposed for modeling MEOR.
\item Conclusion and future work. 
\end{itemize}
\newpage
\section{Reservoir modeling}
Let us consider a porous medium filled with water and oil. We assume that the fluids are immiscible and incompressible. For knowing the amount of a phase in the representative element volume (REV), we introduce the saturation of phase $\alpha$ (for oil $\alpha=o$ and for water $\alpha=w$) given by the ratio of volume of phase $\alpha$ (in REV) over the volume of voids (in REV). In the case where the porous medium is just filled with two fluids, we have that $s_o+s_w=1$.\par 

In the oil-water interface there is a surface free energy due to natural electrical forces, which attract the molecules to the interior of each phase and to the contact surface \cite{Hassanizadeh:Article:1990}. The IFT keeps the fluids separated and it is defined by the quantity of work needed to separate a surface of unit area from both fluids.\par 

Capillary pressure $p_c$ is the difference in pressure between two immiscible phases of fluids occupying similar pores due to IFT between the phases \cite{Bahadori:Book:2014}. It is known that $p_c$ is not a well-defined function because to one value of water saturation, corresponds more than one value of $p_c$, due to $p_c$ being dependent on the history. It means that different saturation values are expected during imbibition as in drainage. This phenomenon occurring in many porous media systems is called hysteresis.\par

We write Darcy's law and the mass conservation equations for each $\alpha$ phase ($\alpha=o,\;w$) 
\begin{equation}\label{tpf}
\frac{\partial (\phi s_\alpha)}{\partial t}+\nabla\cdot\textbf{u}_\alpha=\frac{F_\alpha}{\rho_\alpha}\;\;\;\;\;\;\;\;\;\;\textbf{u}_\alpha=-\lambda_\alpha\textbf{k}(\nabla p_\alpha-\rho_\alpha\textbf{g}),\\ 
\end{equation}
where $\phi$ is the porosity, $\textbf{u}_\alpha$ the volumetric flow rate per area, $F_\alpha$ the source/sink term, $\rho_\alpha$ the density, $\textbf{k}$ the absolute permeability and $\lambda_\alpha=\tfrac{k_{r,\alpha}}{\mu_\alpha}$ the phase mobility, with $k_{r,\alpha}$ the relative permeability and $\mu_\alpha$ the viscosity. In this work we consider that the porosity does not change over time. Defining the average pressure $p=\tfrac{1}{2}(p_w+p_o)$, $\lambda_\Sigma=\lambda_o+\lambda_w$ and $\lambda_\Delta=\lambda_o-\lambda_w$, using $s_w+s_o=1$ and $p_w-p_o=p_c$, we  can reformulate the problem solving for $p$ and $s_w$
\begin{gather}\label{presse}
\begin{split}
&\text{Pressure equation}\;\;\;\;-\nabla\cdot(\textbf{k} (\lambda_\Sigma\nabla p+\frac{1}{2}\lambda_\Delta\nabla p_c-(\lambda_w\rho_w+\lambda_o\rho_o)\textbf{g}))=\sum_{\alpha=w,o}\frac{F_\alpha}{\rho_\alpha}.\\
&\text{Saturation equation}\;\;\;\;\phi\frac{\partial s_w}{\partial t}-\nabla\cdot (\lambda_w\textbf{k}(\nabla (p-\frac{1}{2}p_c)-\rho_w\textbf{g}))=\frac{F_w}{\rho_w}.
\end{split}
\end{gather}
An extended description of the previous equations can be found e.g. in \cite{Nordbotten:Book:2011} and \cite{Helmig:Book:1997}. 
\subsection{IFA}\label{interarea}
\begin{figure}[h!]
\normalsize
\centering
\includegraphics[width=3.9cm,height=1.8cm]{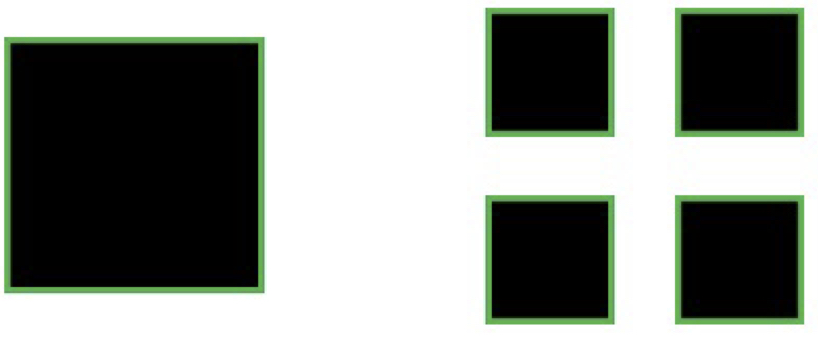}
\caption[]{\raggedright Comparison of IFA given the same amount of oil} 
\label{Fig1}
\end{figure}

Considering a porous medium filled with two fluids, the surface where they make contact is called IFA. Mathematically, we compute the specific IFA $a_{ow}$ as a ratio of the IFA in the REV over the volume of REV. For understanding better the importance of $a_{ow}$ in the oil recovery, let us consider Fig. \ref{Fig1}, where we observe that splitting the square in four pieces, the IFA increases by a factor of 2. Then, we can recover faster the oil in the zones with larger IFA.\par
\newpage
When Darcy made his experiments and deduced his law, he just considered a single-phase flow. In the case of two-phase flow, we just extend Darcy's law for two fluids, but we may expect there are more forces involve than the gradient of the hydraulic head. In \cite{Hassanizadeh:Article:1990} was developed equations of momentum balance for phases and interfaces, based on thermodynamic principles. In addition, equations of balance of mass for phases and interfaces are considered \cite{Hassanizadeh:Article:1993a}. After performing various transformations \cite{Niessner1:Article:2008}, the following balance equation of specific interfacial area for the oil-water interface is obtained
\begin{equation}\label{mamamia}
\frac{\partial a_{ow}}{\partial t}+\nabla\cdot(a_{ow}\textbf{v}_{ow})=E_{ow} \;\;\;\;\;\;\;\text{with}\;\;\;\;\;\;\;\textbf{v}_{ow}=-\textbf{k}_{ow}\nabla a_{ow},
\end{equation}
where $\textbf{v}_{ow}$ is the interfacial velocity, $E_{ow}$ is the rate of production/destruction of specific IFA and $\textbf{k}_{ow}$ is the interfacial permeability. Based on a thermodynamic approach, \cite{Hassanizadeh:Article:1993} demonstrated that including the IFA in the capillary pressure relation reduces the hysteresis under equilibrium conditions. \par

In order to close our model, which includes the specific oil-water IFA, we have to provide a relation $p_o-p_w=p_c(a_{ow},s_w)$ that accounts for interfacial forces. This relation can be obtained by fitting surfaces to $a_{ow}-s_w-p_c$ data coming from models or experiments. In \cite{Niessner1:Article:2008}, they used a bi-quadratic relationship. However, this relation does not fulfill the requirements $a_{ow}(0,p_c)=a_{ow}(1,p_
c)=0$. In this work, we use the next relation \cite{Joekar-Niasar:Article:2012}
\begin{equation}\label{awo}
a_{ow}(s_w,p_c)=\alpha_1 s_w^{\alpha_2}(1-s_w)^{\alpha_3}p_c^{\alpha_4}
\end{equation}
with $\alpha_1$, $\alpha_2$, $\alpha_3$ and $\alpha_4$ constants. From the previous parameterization, we can isolate the capillary pressure
\begin{equation}\label{pca}
p_c(s_w,a_{ow})=\alpha_1^{-1/\alpha_4}s_w^{-\alpha_2/\alpha_4}(1-s_w)^{-\alpha_3/\alpha_4}a_{ow}^{1/\alpha_4}.
\end{equation}
For solving the specific IFA equation, we need to provide the mathematical expression for $E_{ow}$. In \cite{Niessner1:Article:2008} was proposed the following relation based on physical arguments
\begin{equation}
E_{ow}=-e_{ow}\frac{\partial s_w}{\partial t},
\end{equation}
where $e_{ow}$ is a parameter characterizing the strength of change of specific IFA due to a change of saturation
\begin{equation}\label{porque}
e_{ow}=-\frac{\partial a_{ow}}{\partial p_c}\bigg (\frac{d p_c}{ds_w} \bigg )_{\text{line}}-\frac{\partial a_{ow}}{\partial s_w}.
\end{equation}
\newpage
The path $\bigg (\frac{dp_c}{ds_w} \bigg )_{\text{line}}$ is in general unknown, but in the main drainage and imbibition curves, $p_c$ is a known function of $s_w$. In addition, it is possible to compute this derivative for $e_{ow}=0$. For all other paths, we interpolate using these three values of $e_{ow}$ \cite{Niessner1:Article:2008}.\par

Experimental investigations focused on simultaneously measuring $p_c$, $s_w$ and $a_{ow}$ are often difficult, expensive and subject to limitations, thus only a few have been reported in the literature, indicating a need for further experimental studies characterizing the relationship $a_{ow}(s_w,p_c)$ \cite{Chen:Article:2007, Porter:Article:2010}. 
\subsection{Transport equations}
For describing the movement of bacteria, nutrients and surfactants, we consider the following transport equations ($\beta=\lbrace b,n,s \rbrace$)
\begin{equation}
\frac{\partial (C_\beta\phi s_w)}{\partial t}-\nabla\cdot\bigg (\textbf{D}_\beta \phi s_w\nabla C_\beta-\textbf{u}_wC_\beta-\delta_{b\beta}(\textbf{v}_aC_b+\textbf{v}_g\phi C_b)\bigg )=R_\beta,
\end{equation}
where the reaction rate terms are given by \cite{Kim:Article:2006, Li:Article:2011}
\begin{gather}
\begin{split}
R_b=g_{1}\phi s_wC_b-d_1\phi s_wC_b-\frac{R_s}{Y_{s/b}}\;\;\;\;\;\;\;\;R_n=-\frac{R_s}{Y_{s/n}}-Y_n\phi s_w C_b\;\;\;\;\;\;\;\;R_s=\mu_{\text{s}}\phi s_wC_b
\end{split}
\end{gather}
and in the general case the dispersion coefficients are given by
\begin{equation}
D_{\beta,ij}=\delta_{ij}\alpha_{\beta,T} |\textbf{u}|+(\alpha_{\beta,L}-\alpha_{\beta,T})\frac{u_iu_j}{|\textbf{u}|}+\delta_{ij}D_{\;\;\beta}^{\text{eff}},
\end{equation}
where the fluid velocity of the aqueous phase is given by $\textbf{u}=\frac{\textbf{u}_w}{\phi s_w}$. In the previous equations, $C_b$, $C_n$, $C_s$ are the concentrations of bacteria, nutrients and surfactants, $\alpha_{\beta,L}$ the longitudinal dispersivity, $\alpha_{\beta,T}$  the transverse dispersivity, $D_\beta^\text{eff}$ the effective diffusion coefficients of bacteria, nutrients and surfactants in the water phase and $\delta_{ij}$ the Dirac delta. We consider that the bacteria, nutrients and surfactants live on the water, so their transport due to the convection is given by the term $\textbf{u}_wC_\beta$. We include gravity effects on the bacteria considering the settling velocity of bacteria $v_g$. For including that the bacteria has a tendency to live in the oil-water interface \cite{Kosaric:Book:2015}, we add the chemotactic velocity $\textbf{v}_a$ in the bacteria transport equation. We propose the following expression for the chemotactic velocity
\begin{equation}
\textbf{v}_a=k_a\nabla a_{ow},
\end{equation}
where $k_a$ is a diffusive term. It is for the first time when such a chemotaxis term is included in the modeling transport of bacteria in two-phase porous media. Including the chemotaxis in MEOR models is important because besides the external constrains, it also determines the distribution of bacteria in the soil \cite{Gharasoo:Article:2014, Centler:Article:2015}.\par
    
Let us analyses the reaction terms for the transport equations. For modeling the growth of bacteria, we use the Monod-type model \cite{kovarov:Article:1998}
\begin{equation}
g_1=g_{1\text{max}}\frac{C_n}{K_{b/n}+C_n},
\end{equation}
\newpage
\noindent where $g_{1\text{max}}$ is the observed maximum growth rate and $K_{b/n}$ the half saturation constant, being the nutrient concentration level when $g_1=\tfrac{1}{2}g_{1\text{max}}$. On the other hand, we consider a linear death of bacteria, given by $d_1$. Due to nutrients and bacteria being involved in the generation of surfactants, we introduce the surfactant yield coefficients $\frac{1}{Y_{s/b}}+\frac{1}{Y_{s/n}}=1$. For the nutrients consumed for bacteria, we consider the yield coefficient $Y_n$, which we included in the $R_n$ term. In the absence of IFA, one relation for the production rate of surfactants is given by \cite{Lacerda:Article:2012}
\begin{equation}
\mu_s(C_n)=\mu_{\text{s max}}\frac{C_n-C_n^*}{K_{s/n}+C_n-C_n^*},
\end{equation}
where $\mu_{\text{s max}}$ is the maximum specific biomass production rate, and $C_n^*$ the critical nutrient concentration for metabolism term, that models a need of minimum $C_n$ for obtaining surfactants. One of the characteristics that a surfactant should have is biological production at the oil-water interface \cite{Donaldson:Book:1989}. In order to consider this effect in our model, we consider the production rate of surfactant as a function of the nutrient concentration and IFA. To our knowledge, there are not experimental studies to deduce a mathematical relation of the surfactant production in function of the IFA; therefore, we need experiments for $\mu(C_n,a_{ow})$. Given the mathematical characteristics of the Monod-type function, we propose the following expression for the production rate of surfactants
\begin{equation}
\mu_s(C_n,a_{ow})=\mu_{\text{s max}}\frac{a_{ow}}{K_{a}+a_{ow}}\frac{C_n-C_n^*}{K_{s/n}+C_n-C_n^*},
\end{equation}
where $K_a$ is the half saturation constant.\par
The pressure, saturation and IFA equations are coupled with these transport equations under the assumptions that the two-phase flows are incompressible and immiscible, both viscosities are constants, the presence of dissolved salt in the wetting phase is neglected and the system is isothermal \cite{Li:Article:2011}.
\subsection{IFT}
One of the main objectives of applying MEOR is to reduce the $s_{or}$ via surfactant effect on the oil-water IFT. There exist several experiments showing the impact of surfactants in reducing the IFT \cite{Wu:Article:2014}. Common initial IFT values are of the order of $10^{-2}$ mN/m and we aim to lower this value $\leq\;10^{-3}$ mN/m \cite{Yuan:Article:2015}. One mathematical model for the IFT reduction is given by \cite{Nielsen:Article:2015}
\begin{equation}\label{mamama}
\sigma=\sigma_0\frac{-\tanh(l_3C_{s}-l_2)+1+l_1}{-\tanh(-l_2)+1+l_1},
\end{equation}
where $\sigma_0$ is the initial IFT, $l_1,\;l_2$ and $l_3$ are fitting parameters, which define the efficiency of the surfactant, moderating the concentration where the IFT drops dramatically and the minimal IFT achieved after the surfactant action \cite{Nielsen:Article:2015}.\par
When the surfactant concentration increases, the IFT and $p_c$ decrease. For considering this effect in our model, we include the dependence of the $C_s$ in Eq. \ref{pca}, resulting in the following capillary pressure expression
\begin{equation}\label{pepo}
p_{c}(s_w,a_{ow},C_s,\phi,\textbf{k})=\sigma(C_s) \sqrt{\tfrac{\phi}{||\textbf{k}||}}\alpha_1^{-1/\alpha_4}s_w^{-\alpha_2/\alpha_4}(1-s_w)^{-\alpha_3/\alpha_4}a_{ow}^{1/\alpha_4}
\end{equation}
where we also include the porosity and permeability. Then, the IFA becomes
\begin{equation}\label{chiquita}
a_{ow}(s_w,p_c,C_s,\phi,\textbf{k})=\alpha_1 s_w^{\alpha_2}(1-s_w)^{\alpha_3}\bigg (\frac{p_c}{\sigma(C_s) }\sqrt{\frac{||\textbf{k}||}{\phi}}\bigg)^{\alpha_4}.
\end{equation}
\subsection{Trapping number}
The residual oil saturation after water flooding is believed to be distributed through the pores in the petroleum reservoir in the form of immobile globules, being the capillary and viscous interactions the main forces acting on these globules \cite{Donaldson:Book:1989}. The capillary number $N_{\text{Ca}}$ relates the surface tension and viscous forces acting in the interface, the bond number relates the buoyancy to capillary forces and the trapping number $N_\text{T}$ quantifies the force balance. Then, the mathematical expressions for these numbers are given by \cite{Pennell:Article:1996}
\begin{equation}\label{trap}
N_{\text{Ca}}=\frac{u_w\mu_w}{\sigma\cos\theta}\;\;\;\;\;\;\;\;\;\; N_\text{B}=\frac{(\rho_w-\rho_n) g k k_{r,w}}{\sigma\cos\theta}\;\;\;\;\;\;\;\;\;\; N_\text{T}=\sqrt{N^2_{\text{Ca}}+2N_{C_a}N_{B}\sin\alpha_0+N^2_\text{B}},
\end{equation}
where $\theta$ is the contact angle between the oil-water interface and $\alpha_0$ is the angle of flow relative to the horizontal. At the end of water flooding, the capillary number is in the range $10^{-6}$ to $10^{-7}$ \cite{Donaldson:Book:1989}. In order to increase the capillary number, from Eq. \ref{trap} we observe that increasing the flow rate, the water viscosity or lowering the IFT are the three possibilities. In \cite{Li:Article:2011}, they stated that MEOR could improve the oil extraction if we can obtain a capillary number between $10^{-5}$ and $10^{-1}$.
\subsection{Residual oil saturation}
For relating the residual oil saturation and the capillary number, we use the following relation (\cite{Li:Article:2007})
\begin{equation}\label{sorsor}
s_{or}=\min \big (s_{or},s_{or}^{\text{min}}+(s_{or}^{\text{max}}-s_{or}^{\text{min}})[1+(T_1N_{\text{T}})^{T_2}]^{\tfrac{1}{T_2}-1} \big ),
\end{equation}
where $s_{or}^{\text{min}}$ and $s_{or}^{\text{max}}$ are the maximum and minimum residual oil saturation and both $T_1$  and $T_2$ are fitting parameters estimated from the experimental data.\par
Giving the mathematical expressions for the IFT reduction, the trapping number and the residual oil saturation reduction, we can account in the model the effect of the surfactants in improving the oil recovery.
\subsection{Two-phase flow model with transport equations including IFA}
In summary, we propose the next set of equations as the first complete model for MEOR including IFA effects
\newpage
\begin{flushleft}
\textbf{Pressure}\\
$-\nabla\cdot(\textbf{k} (\lambda_\Sigma\nabla p+\frac{1}{2}\lambda_\Delta\nabla p_c-(\lambda_w\rho_w+\lambda_o\rho_o)\textbf{g}))=\sum\limits_{\alpha=w,n}\frac{F_\alpha}{\rho_\alpha}$\\
\textbf{Saturation}\\
$ \phi \frac{\partial s_w}{\partial t}-\nabla\cdot (\lambda_w\textbf{k}(\nabla (p-\frac{1}{2}p_c)-\rho_o\textbf{g}))=\frac{F_w}{\rho_w}$\\
\textbf{Interfacial area}\\
$\frac{\partial a_{ow}}{\partial t}-\nabla\cdot(a_{ow}\textbf{k}_{ow}\nabla a_{ow})=(\frac{\partial a_{ow}}{\partial p_c} (\frac{d p_c}{ds_w} )_{\text{line}}+\frac{\partial a_{ow}}{\partial s_w})\frac{\partial s_w}{\partial t}$\\
\textbf{Bacterial concentration}\\
$\frac{\partial (C_b\phi s_w)}{\partial t}-\nabla\cdot (\textbf{D}_b \phi s_w\nabla C_b-\textbf{u}_wC_b-k_a\nabla a_{o,w}C_b-\textbf{v}_g\phi C_b)=(g_{1\text{max}}\frac{C_n}{K_{b/n}+C_n}-d_1)\phi s_wC_b-\frac{R_s}{Y_{s/b}}$\\
\textbf{Nutrient concentration}\\
$\frac{\partial (C_n\phi s_w)}{\partial t}-\nabla\cdot (\textbf{D}_n \phi s_w\nabla C_n-\textbf{u}_wC_n )=-\frac{R_s}{Y_{s/n}}-Y_n\phi s_w C_b$\\
\textbf{Surfactant concentration}\\
$\frac{\partial (C_s\phi s_w)}{\partial t}-\nabla\cdot (\textbf{D}_s \phi s_w\nabla C_s-\textbf{u}_wC_s )=\mu_{\text{s max}}\frac{a_{ow}}{K_{a}+a_{ow}}\frac{C_n-C_n^*}{K_{s/n}+C_n-C_n^*}\phi s_wC_b$\\
\textbf{Relative permeabilities}\\
$k_{r,w}(s_w)=(\frac{s_w-s_{wi}}{1-s_{or}-s_{wi}})^2$ $k_{r,o}(s_w)=(\frac{1-s_w-s_{or}}{1-s_{or}-s_{wi}})^2$\\
\textbf{Capillary pressure}\\
$p_{c}(s_w,a_{ow},C_s,\phi,||\textbf{k}||)=\sigma(C_s) \sqrt{\tfrac{\phi}{||\textbf{k}||}}\alpha_1^{-1/\alpha_4}s_w^{-\alpha_2/\alpha_4}(1-s_w)^{-\alpha_3/\alpha_4}a_{ow}^{1/\alpha_4}$\\
\textbf{Interfacial tension}\\
$\sigma=\sigma_0\frac{-\tanh(l_3*C_{s}-l_2)+1+l_1}{-\tanh(-l_2)+1+l_1}$\\
\textbf{Trapping number}\\
$N_{\text{Ca}}=\frac{u_w\mu_w}{\sigma\cos\theta}\;\;\;\;\;\;\;\;\;\;\;\;\;\;\;\;\;\;\;\; N_\text{B}=\frac{(\rho_w-\rho_n) g k k_{r,w}}{\sigma\cos\theta}\;\;\;\;\;\;\;\;\;\;\;\;\;\;\;\;\;\;\;\; N_\text{T}=\sqrt{N^2_{\text{Ca}}+2N_{C_a}N_{B}\sin\alpha_0+N^2_\text{B}}$\\
\textbf{Residual oil saturation}\\
$s_{or}=\min \big (s_{or},s_{or}^{\text{min}}+(s_{or}^{\text{max}}-s_{or}^{\text{min}})[1+(T_1N_{T})^{T_2}]^{\tfrac{1}{T_2}-1} \big )$.
\end{flushleft}
\section{Discretization and implementation}
\begin{figure}[h!]
\centering
\includegraphics[scale=.5]{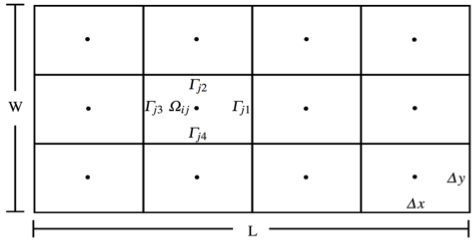}
\caption{A uniform cell-centered grid in a rectangular domain of size $L\;\times\;W$}
\label{Fig2}
\end{figure}

After having set the model equations, we proceed to define the space domain. We consider a rectangular domain with a uniform cell-centered grid with half-cells at the boundaries. Fig. \ref{Fig2} shows a uniform cell-centered grid in a space domain of length $L$ and width $W$.\\
\\
Discretization of time is achieved considering a uniform partition from the initial time $t_0=0$ until the final time $T$ with $\Delta t$ the time step. After discretizing the space and time, now we discretize the derivatives and integrals. Considering an arbitrary function $f(x,y)$, using its Taylor expansion we get the following approximation for the derivatives
\begin{equation}
\frac{\partial f}{\partial x}=\frac{f(x+\Delta x,y)-f(x,y)}{\Delta x}+O(\Delta x)\;\;\;\;\;\;\;\;\frac{\partial f}{\partial x}=\frac{f(x+\tfrac{\Delta x}{2})-f(x-\tfrac{\Delta x}{2})}{\Delta x}+O(\Delta x^2).
\end{equation}
The approximation of first order is used on the boundaries of the spatial domain and time derivatives while the second order approximation is used in the cell-centered grill. To discretize time derivatives $u_t(x,t)=F(u,t)$, we consider the BE method
\begin{equation}
u^{n+1}_{i}=u_i^n+F(u_i^{n+1},t^{n+1})\Delta t+O(\Delta t).
\end{equation}
Finally, for approximating integrals, we use the midpoint rule
\begin{equation}
\int\limits_{\Omega_{ij}}f(x,y)dxdy\approx\Delta x\Delta yf_{ij}.
\end{equation}   
In the previous section we developed a two-phase flow model for MEOR with transport equations including IFA effects. We are interested in the solution of this system. In order to perform numerical simulations, it is necessary to discretize these equations. In this work, we used a cell-centered finite-volume method called  TPFA. A detailed description about how to use and implement TPFA in MATLAB can be found in \cite{Aarnes:Book:2007}. As we consider a cell centered grid, we do not know the values of the parameters on the walls, so it is necessary to use an approximation on the boundaries. Depending on the parameter, we should consider different technique approximations, in order to get stability and correct results \cite{Aziz:Article:1979}. Regarding the permeability of the medium, we approximate $k$ by the harmonic mean \cite{Aavatsmark:Article:2002}
\begin{equation}\label{permeability}
k_{i+1/2}=2(\Delta x_{i+1}+\Delta x_i)\bigg (\frac{\Delta x_{i+1}}{k_{i+1}}+\frac{\Delta x_i}{k_{i}}\bigg )^{-1}.
\end{equation}
The reason for considering this harmonic mean comes from the computation of an effective permeability when we consider a layered system with different values of permeability and a flux perpendicular to these layers, finding that the effective permeability of a system with two layers is given by the previous equation \cite{Nordbotten:Book:2011}. For the rest of the parameters that we need to approximate on the walls, we simply use the average value
\begin{equation}
\xi_{i+1/2}=\frac{\xi_{i+1}+\xi_{i}}{2}.
\end{equation}
There are several algorithms to solve reactive transport models \cite{Li:Article:2011, vanWijngaarden:Article:2011, Nick:Article:2013}. In this work, for solving the pressure, saturation and IFA equations, we use an implicit scheme. The use of these iterative formulations is very common, for example in \cite{Pop:Article:2004} and \cite{List:Article:2016} they solved the Richards equation using this technique. Regarding the two-phase flow, in \cite{Radu:Article:2015} they solved the pressure and saturation equations using the same iterative scheme. The convergence of this implicit scheme can be followed from  \cite{Radu:Article:2010}, \cite{Kumar:Article:2013} and \cite{Kumar:Article:2014}.\par
Due to the capillary pressure is a function of the saturation and interfacial area, both of them being unknowns, we use an inner iteration ${}^j$ in order to upgrade the values of the functions depending on the saturation and interfacial area and solve this system of equations (pressure, saturation and IFA equations) until a stopping criterion is reached. For initializing the iteration, we consider the solution at the previous time step
\begin{equation}
p_i^{n+1,1}=p_i^n,\;\;\;a_{wn,i}^{n+1,1}=a_{wn,i}^n,\;\;\;s_{w,i}^{n+1,1}=s_{w,i}^n,\;\;\;\lambda_i^{n+1,1}=\lambda_i^n,\;\;\;\forall i.
\end{equation}
When we discretized the pressure and saturation equation, we used the chain rule for computing the gradient of the capillary pressure 
\begin{equation}
\nabla p^{n+1,j+1}_{c,ij}=\partial_{s_w} p^{n+1,j}_c(s^{n+1,j}_{w,ij},a^{n+1,j}_{ow,ij})\nabla{s}^{n+1,j+1}_{w,ij}+\partial_{a_{ow}} p_c^{n+1,j}(s^{n+1,j}_{w,ij},a^{n+1,j}_{ow,ij})\nabla{a}^{n+1,j}_{ow,ij}
\end{equation}
in order to improve the stability of the scheme.\par 

To solve the transport equations, we use an iterative solver
\begin{gather}
\begin{split}
\frac{C^{n+1,j+1}_\beta\phi s^{n+1}_w-C^{n}_\beta\phi s^{n}_w}{\Delta t}&-\nabla\cdot\bigg (\textbf{D}^{n+1}_\beta \nabla \phi s^{n+1}_w C^{n+1,j+1}_\beta-\textbf{u}^{n+1}_wC^{n+1,j+1}_\beta\\
&-\delta_{b\beta}(k_a\nabla a^{n+1}_{ow}C^{n+1,j+1}_b+\textbf{v}_g\phi C^{n+1,j+1}_b)\bigg )=R^{n+1,j}_\beta.
\end{split}
\end{gather}
We write the three system of equations in the same matrix, looking for the solution of $(C^{n+1,j+1}_{b,1},\;C^{n+1,j+1}_{n,1},\;C^{n+1,j+1}_{p,1},\;\ldots,C^{n+1,j+1}_{p,N})^T$ iteratively until a stopping criterion is reached. \par

Due to we use an iterative scheme, it is necessary to have a measurement of the error. For this work, we use the following $L^2$-norm 
\begin{equation}
\|\textbf{x}\|_{L^2}=\bigg (\Delta x\sum\limits_{i=1}^m x_i^2\bigg )^{1/2}.
\end{equation}
Then, the algorithm for solving the model equations is the following
\begin{enumerate}
\item We solve the pressure equation using the previous values of water saturation and IFA.
\item We solve the saturation equation using the updated values of pressure but the previous values of IFA.
\item We solve the IFA equation using the updated values of saturation.
\item We compute the errors $\|{}^{j+1}\textbf{p}^{n+1}-{}^j\textbf{p}^{n+1} \|_{L^2}$, $\|{}^{j+1}\textbf{s}^{n+1}-{}^j\textbf{s}^{n+1} \|_{L^2}$ and $\|{}^{j+1}\textbf{a}^{n+1}-{}^j\textbf{a}^{n+1} \|_{L^2}$.
\item If the errors are less than a given tolerance $\epsilon$, we proceed to solve the concentration equations. Otherwise, we upgrade the values for the inner iteration ${}^j$ and we solve again the three equations. If any of the errors does not get less than $\epsilon$ in a given maximum number of iterations $MI_n$, we halve the time step and try again. If we halve the time step in a maximum number of times $MI_{\epsilon}$, we have to check if the problem is well-posed. 
\item We solve the concentration equations iteratively until the error is less than $\epsilon$ or we proceed as mentioned before halving the time step.
\item If the concentration error is less than $\epsilon$, we compute the IFT, $N_T$ and $s_{\text{sor}}$.
\item We move to the next time step and we repeat the process until we reach the final time T and we plot the results.
\end{enumerate}  
\newpage
\section{Results and discussion}
Following all previous work, we can finally perform numerical experiments to study the effects of MEOR considering the oil-water IFA. In order to formulate the model, we considered the next works: \cite{Li:Article:2011} (Transport equations), 
\cite{Niessner1:Article:2008} and \cite{Joekar-Niasar:Article:2012} (IFA), 
\cite{Nielsen:Article:2015} (reduction of IFT) and
\cite{Li:Article:2007} (reduction of $s_{or}$).\par

Using the best estimate of physical parameters from the existing experiments, we were not able to obtain physically plausible results. We interpret this to be due to the disparate experimental conditions used in the cited works, leading to results which are physically incompatible. Thus the first conclusion of our work is that the existing experimental literature for MEOR and interfacial area is incomplete, and that dedicated experiments encompassing the full process of microbial growth, transport and surfactant production together with changing IFA, new relations for the rate of production/destruction of IFA and new capillary pressure surfaces are needed.\par
In lieu of complete and compatible experimental data, we have thus conducted numerical simulations with what we deem plausible data, to highlight the dominating physical processes in the system.\\
We consider a porous medium of length $L=2$ m and width $W=1$ m. We set the initial water saturation as $s_w(x,y,0)=0.3+0.4*y$. We inject water, bacteria and nutrients into the left boundary and oil, water, bacteria, nutrients and surfactants flow out through the right boundary. There is not flux through the upper and bottom boundary. For the water and oil pressures, we take the same conditions as in \cite{Li:Article:2011}: $p_w(x,y,0)=0.981$ kPa and $p_o(x,y,0)=9.417$ kPa; leading to an average pressure of $p(x,y,0)=5.199$ kPa and initial capillary pressure of $p_c(x,y,0)=8.436$ kPa. On the left boundary, we have a flux boundary condition $Q_T/A=-2.78\times 10^{-5}\text{m}\;\text{s}^{-1}$. Due to we inject water, the left boundary condition for the water saturation is $s_w(0,y,t)=0.7$. Regarding the right boundary condition for the water saturation, we consider a Neumann condition with zero value. We choose the initial value and left boundary of IFA evaluating Eq. \ref{chiquita} with the initial and left values of water saturation, capillary pressure, IFT, permeability and porosity respectively. We consider that there is neither bacteria nor nutrients initially in the porous media and we inject them on the left boundary with a concentration of $0.5\;kg\;m^3$. We also consider a no-flux boundary condition for the surfactant concentration on the left boundary. Regarding the relation $p_c(s_w)$ in Eq. \ref{porque}, we use Eq. \ref{pepo} evaluating with the initial average IFA, IFT, permeability and porosity. Table \ref{tab:1} shows the value parameters used in the numerical simulations.
\begin{table}[h!]
\centering
\caption[]{\raggedright Parameters}
\small{
\begin{tabular}{ll}
  \hline\noalign{\smallskip}			
  Parameter$\;\;\;\;\;\;\;\;\;\;\;\;\;\;\;\;\;\;\;\;\;\;\;\;\;\;\;\;\;\;\;\;\;\;\;\;\;\;\;\;\;\;\;\;\;\;\;\;\;\;\;\;\;\;\;\;\;$& Value\\
  \noalign{\smallskip}\hline\noalign{\smallskip}
  $\phi$ & 	$0.4$\\
  $k$ &	$0.94\;(\times 10^{-12})\text{m}^2$\\
  $\mu_w$	&	$1\times 10^{-3}\;\text{kg}\;\text{m}^{-1}\;\text{s}^{-1}$\\
  $\mu_o$	&	$3.92\times 10^{-3}\;\text{kg}\;\text{m}^{-1}\;\text{s}^{-1}$\\
  $\rho_w$	&	$1000\;\text{kg}\;\text{m}^{-3}$	\\
  $\rho_o$	&	$800\;\text{kg}\;\text{m}^{-3}$	\\
  $K_a$	&	$25\;\text{m}^{-1}$\\
  $g$	&	$0$\\
  $v_g$ & $0$ \\
  $g_{1\;\text{max}}$	&	$5\times 10^{-4}\text{s}^{-1}$\\
  $d_1$	&	$5\times 10^{-4}\text{s}^{-1}$\\
  $\mu_{p\;\text{max}}$	& $1\times 10^{-4}\text{s}^{-1}$\\
  $C_n^{*}$	&	$0$\\	
  $K_{b/n}$	&	$1$ kg m${}^{-3}$\\
  $K_{s/n}$	&	$1$ kg m${}^{-3}$\\
  $Y_{s/b}$	&	$1$\\
  $Y_{s/n}$	&	$9$\\ 
  $Y_{n}$	&	$2\times 10^{-4}\text{s}^-1$\\
  $\alpha_{b,T}$, $\alpha_{n,T}$, $\alpha_{s,T}$	&	$0.01\;\text{m}$\\
  $D_{b}^{\text{eff}}$, $D_{n}^{\text{eff}}$, $D_{s}^{\text{eff}}$	&	$1.5\times 10^{-9}\;\text{m}^2\;\text{s}^{-1}$ \\
  $\alpha_1$& 279.56\\
  $\alpha_2$& 0\\
  $\alpha_3$& 1.244\\
  $\alpha_4$& -0.963\\
  $k_{ow}$	&	$1\times 10^{-8}\;\text{m}^3\text{s}^{-1}$ \\
  $k_{a}$	&	$1\times 10^{-8}\;\text{m}^3\text{s}^{-1}$ \\
  $l_1$&$41\times 10^{-4}$\\
  $l_2$&$12$\\
  $l_3$&$180$\\
  $T_1$&$5\times 10^{4}$\\
  $T_2$&$10$\\
  $s_{wi}$	&	$0.2$\\
  $s_{or}^{max}$&0.3\\
  $s_{or}^{min}$&0.08\\
  $\sigma_0$&$3.37\times 10^{-3}\;\text{N}\;\text{m}^{-1}$\\
  \noalign{\smallskip}\hline
\end{tabular}
}
\label{tab:1}
\end{table}
\newpage
\subsection{IFA effects}
\begin{figure}[h!]
\normalsize
\centering
\includegraphics[scale=.48]{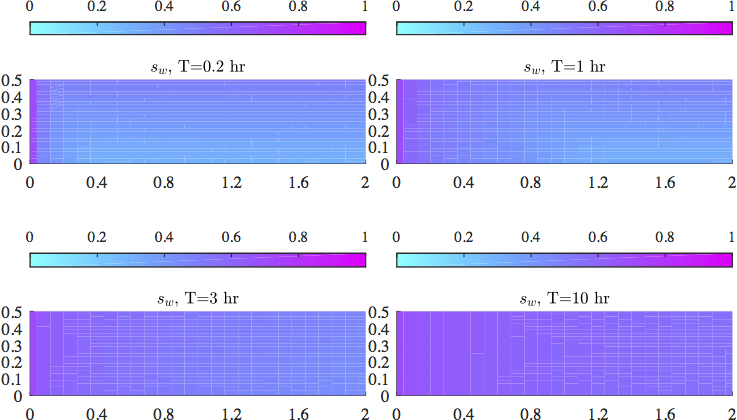}
\caption[]{\raggedright Spatial distribution of the water saturation at different times} 
\label{Fig3}
\end{figure}

\begin{figure}[h!]
\normalsize
\centering
\includegraphics[scale=.48]{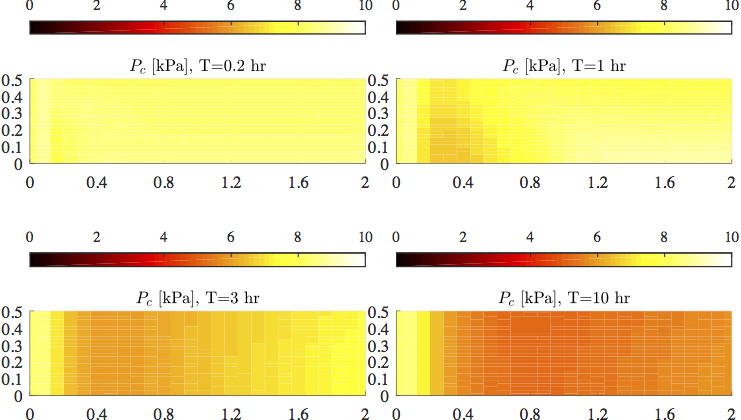}
\caption[]{\raggedright Spatial distribution of the capillary pressure at different times} 
\label{Fig4}
\end{figure}

\begin{figure}[h!]
\normalsize
\centering
\includegraphics[scale=.48]{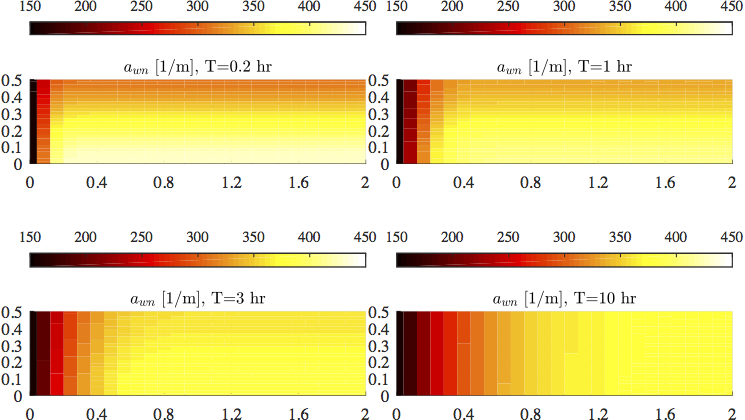}
\caption[]{\raggedright Spatial distribution of the IFA at different times} 
\label{Fig5}
\end{figure}

First, we focus on the IFA effects using the parameters in Table \ref{tab:1}. Fig. \ref{Fig3} shows the evolution in time of water saturation. We notice that in the beginning the upper part has greater water saturation but over time the water saturation in the whole porous media is approaching to the entry saturation.
Fig. \ref{Fig4} shows the evolution in time of capillary pressure. The capillary pressure presents an expected behavior where it is a decreasing function of the water saturation.\\
\newpage
Fig. \ref{Fig5} shows the evolution in time of IFA. We notice that the IFA decreases over time, due to the increment of water saturation.\\
\newpage
\subsubsection{Chemotaxis}
\begin{figure}[h!]
\normalsize
\centering
\includegraphics[scale=.48]{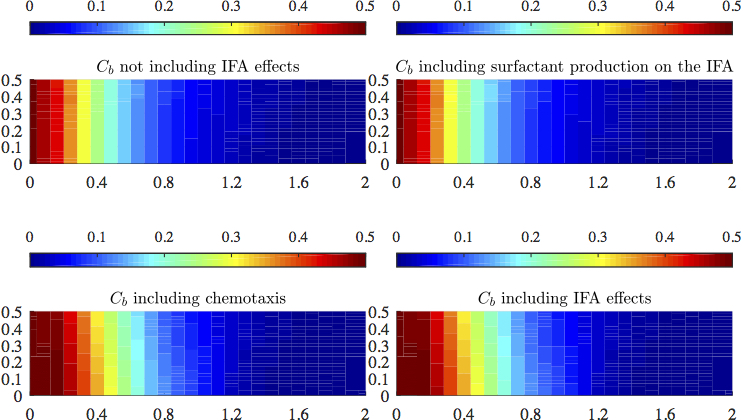}
\caption[]{\raggedright Comparison of the bacterial concentration} 
\label{Fig6}
\end{figure}

In this work we have introduced the term $k_aC_b\nabla a_{ow}$ in the bacteria transport equation in order to model the tendency of surfactants to live in the oil-water interface. To our knowledge, there are not experimental measurements of the coefficient $k_a$. Then, after simulations, we set the value $k_a= 1\times 10^{-8} m^3/s$. Fig. \ref{Fig6} shows the bacterial concentrations in the reservoir after 3.5 hours of water injection for different scenarios. 
\\
From Fig. \ref{Fig5}, we observe that the IFA is increasing from left to right and is greater in the lower part, then we expect to have greater bacterial concentration on the lower part when we consider the chemotaxis; result that we can observe from Fig. \ref{Fig6}.
\subsubsection{Production of surfactants}
\begin{figure}[h!]
\normalsize
\centering
\includegraphics[scale=.48]{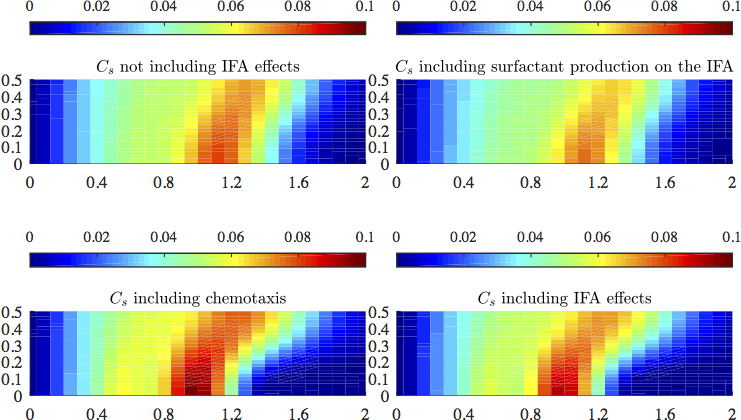}
\caption[]{\raggedright Comparison of the surfactant concentration} 
\label{Fig7}
\end{figure}

In this work, we have also introduced a Monod-type term in the surfactant production rate in order to model the surfactant production at the oil-water interface. Fig. \ref{Fig7} shows the surfactant concentrations in the reservoir after 3.5 hours of water injection for the different scenarios. We observe less surfactant production when we included the Monod-type term because the IFA is increasing from left to right, so when we do not include the surfactant production on the IFA, the surfactant production is overestimated. 
\subsection{MEOR effects}
\begin{figure}[h!]
\normalsize
\centering
\includegraphics[scale=.48]{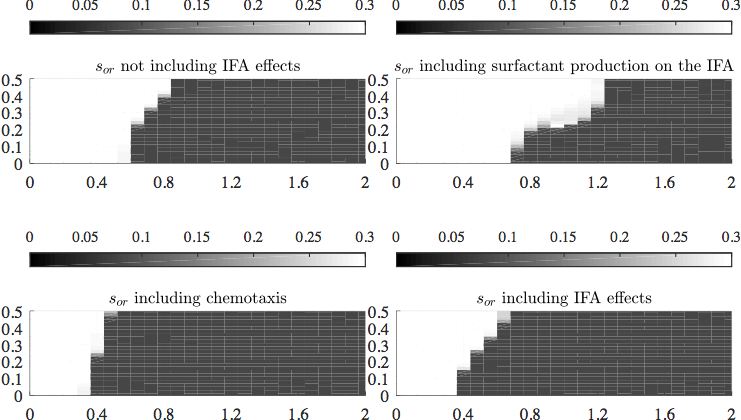}
\caption[]{\raggedright Comparison of the residual oil saturation profiles in the porous media} 
\label{Fig8}
\end{figure}

\begin{figure}[h!]
\normalsize
\centering
\includegraphics[scale=.37]{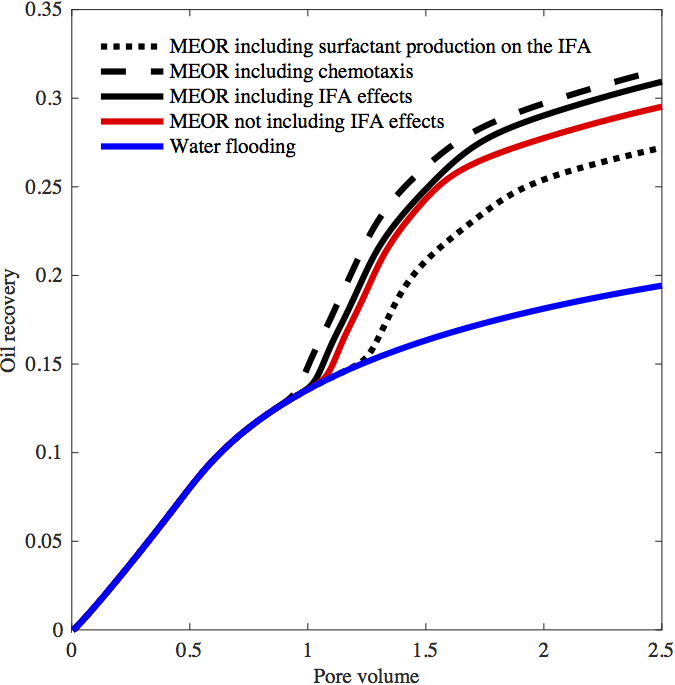}
\caption[]{\raggedright Comparison of the oil recovery in the porous media} 
\label{Fig9}
\end{figure}

The main goal of MEOR it is to enhance the oil recovery using bacteria. Fig. \ref{Fig8} shows the residual oil saturation profiles after 10 hours of injection for the different scenarios.\\
When we just include chemotaxis, we observe the greatest recovery of residual oil saturation, due to the bacteria moves to the zones with greater IFA. When we just included the surfactant production on the IFA, we observe the lowest recovery of residual oil saturation, due to the surfactant production is now limited by the IFA. When we combine both effects, we obtain a greater recovery of residual oil saturation than in the case where we do not include the IFA effects. 
In order to have a measure of the improvements in the oil extraction, we compute the oil recovery. Fig. \ref{Fig9} shows the oil recovery as a function of the pore volume injected in the reservoir. We notice that 10 hours of water injection equals to 2.5 pore volumes. We observe that after injecting approximately 1 pore volumes of water, the surfactant starts to lower the interfacial tension and raise the oil production. When we include the production of surfactants on the IFA, we observe a delay effect in the oil recovery. This delay is due to the production of surfactants is also determined  for the IFA that is increasing from left to right. Then, the rate production of surfactants is less than in the case we do not include the surfactant production on the IFA. When we include the chemotaxis, we observe a faster effect of the surfactants, due to faster migration of bacteria in the reservoir. When we include both chemotaxis and production of surfactant on the IFA, the oil recovery is between these two profiles, being greater than the profile not including the IFA effects.
\newpage
The election of the parameters in Table 1 determined all previous results. It is necessary to estimate all these parameters in the laboratory in order to corroborate the model assumptions. These numerical examples give a better understanding of the mechanisms involve in MEOR.
\section{Conclusions and future perspectives}
A new, comprehensive model for MEOR, which includes two-phase flow, bacteria, nutrient and surfactant transport and considers the role of the oil-water IFA, chemotaxis and reduction of residual oil saturation due to the action of surfactants has been developed. The model particularly includes the oil-water IFA in order to reduce the hysteresis in the capillary pressure relationship, to include the effects of observed bacteria migration towards the oil-water interface and biological production of surfactants at the oil-water interface. To our knowledge, the present work is the first study concerning these effects in the context of MEOR. In particular, the first time to consider the oil-water IFA and chemotaxis for MEOR.\par

The MEOR model consists on a system of nonlinear coupled PDEs and ODEs, whose solution represents a challenge by itself. In order to have an efficient and stable scheme, we used an implicit stepping that considers a linear approximation of the capillary pressure gradient. The time discretization of the equations was obtained using BE and the spatial discretization using FD and TPFA.\par

In order to model that surfactants are produce at the oil-water interface, we considered the production rate of surfactants as a function of the nutrient concentration and IFA in the form of a Monod-type function. To include the chemotaxis, we added the gradient of the IFA in the transport equation for the bacteria.\par

We obtained different water flux profiles and oil recovery predictions when we considered the IFA in the model. In the numerical experiments, we observed an improvement in the oil recovery when we included the IFA effects. Even though real reservoirs are more complex than the model presented, this work is useful for understanding the main phenomena involved in the recovery of petroleum. Moreover, for further calibrating of the present MEOR model, it is necessary to perform more experiments in the laboratory. Through our model, we hope to convince the community for the importance of including IFA and chemotaxis in simulation of MEOR and to inspire further experiments focusing on these relevant effects.\par

Finally, we propose further work inspired in this work. We solved the equations for the pressure, saturation and IFA iteratively, verifying the convergence rate numerically. Nevertheless, it is necessary to do a theoretical analysis of the convergence of the scheme in order to determinate the maximum time step size for having convergence. In order to have a more complete model, we should extend it considering more phenomena, for example bioclogging, surfactant transportation in the oil phase and changes in the viscosities. It is necessary to investigate new relations for the production/destruction rate of IFA $E_{ow}$ because currently there is just one model based on physical arguments but not experimental results.

\subsection*{\small{Acknowledgements}}
\small{This worked was supported by the Research Council of Norway under the project IMMENS no. 255426.}

\end{document}